\documentstyle[sprocl,epsfig]{article}

\newcommand{\gsim}{\;\raisebox{-0.9ex}
                   {$\textstyle\stackrel{\textstyle>}{\sim}$} \;}
\newcommand{\lsim}{\;\raisebox{-0.9ex}
                   {$\textstyle\stackrel{\textstyle<}{\sim}$} \;}

\def\bp {{ \mathbf{p} }}
\def\ls {{\ell s}}
\def\ot {{\otimes}}

\begin{document}
\begin{flushright}
HEPHY--PUB 700/98 \\
STPHY 30/98
\end{flushright}

\vspace{2cm}

\title{MULTIPLICITY DEPENDENCE OF LIKE-SIGN
       AND OPPOSITE-SIGN CORRELATIONS IN $\bar{p}p$ REACTIONS}

\author{B.\ BUSCHBECK$^*$, H.C.\ EGGERS$^\dag$ and P.\ LIPA$^*$\\
\em{$^*$Institut f\"ur Hochenergiephysik, Nikolsdorfergasse 18,
         A--1050 Vienna, Austria\\
         $^\dag$Department of Physics, University of Stellenbosch, 
         7600 Stellenbosch, South Africa}}

\maketitle
\vspace{1cm}

\abstracts{
{\begin{center}{\bf Abstract}\end{center}}
Discussions about Bose-Einstein correlations between decay 
products of coproduced W-bosons again raise the question 
about the behaviour of correlations if several strings are produced. 
This is studied by the multiplicity dependence of correlation functions of particle pairs with like-sign and 
opposite-sign charge in $\bar{p}p$ reactions at $\sqrt{s} = 630$~GeV.}

\vfill

\hrule
To be published in Proceedings of the 8$^{th}$ Internat. Workshop on 
Multiparticle Production ``Correlations and Fluctuations'' in 
M\'atrah\'aza, Hungary, Eds.: T. Cs\"org\"o, S. Hegyi, R.C. Hwa and 
G. Jancs\'o, World Scientific.

\newpage

\section{Introduction}
Recently, there has been much discussion regarding the possibility that 
Bose-Einstein correlations and other interconnection effects~\cite{1}
between decay products of different strings could 
affect the measurement of the $W$ mass. Since measurements are hampered 
by low statistics and experimental difficulties, the question arises whether and
where effects of the superposition of several strings can be tested
independently.
Within the Dual Parton Model, the number of strings may be expected to 
be proportional to multiplicity and thus
influence the multiplicity dependence of various effects, in particular
Bose-Einstein correlations.
While the decrease in observed $\lambda$ as function of multiplicity
can be explained naturally in terms of products of different
strings,~\cite{5,6} other features such as the radius cannot.~\cite{7}

Improvements in experimental analysis techniques,~\cite{8} larger
data samples and the above theoretical background make
it desirable to repeat and extend Bose-Einstein analyses with an
advanced strategy. In this contribution, we investigate 
like-sign and opposite-sign correlations at different total 
multiplicities with the same model-independent strategy and 
good statistics. The bias introduced by selecting events of a given overall 
multiplicity is eliminated with the use of ``internal cumulants''.~\cite{8}

\section{Data sample and normalized density correlation functions}
The data sample consists of 600.000 non-single-diffractive 
$\bar{p}p$ reactions at $\sqrt{s} = 630$~GeV measured by
the UA1 central detector.~\cite{9} As before, only vertex-associated charged 
tracks with transverse momentum $p_{T} \geq 0.15$~GeV/c, $|\eta | 
\leq 3$, good measurement quality and fitted length $\geq$ 30cm have 
been used. To avoid acceptance problems, we restricted the azimuthal 
angle to $45^{0} \leq |\phi | \leq 135^{0}$ (``good azimuth'').

Fig.\ 1 shows the normalized density correlation functions for 
pairs of like-sign charge and for opposite-sign charge separately. Restricting the 
total uncorrected charged multiplicity $N$ in $|\eta | \leq 3$ to 
the windows $1 {\leq} N \leq 6$ (Fig.\ 1a) 
and $28 {\leq} N \leq 35$ (Fig.\ 1b), one 
obtains for the corrected particle density in the central rapidity 
region $dn/d\eta = 0.83 \pm 0.08$ and $dn/d\eta = 5.3 \pm 0.17$ 
respectively\footnote{Events are selected according to multiplicity in {\it all azimuthal\/}
regions, while the subsequent analysis is performed for particles
in the {\it good azimuth\/} only. This is done because the total rather
than the good-azimuth multiplicity is physically relevant.
Total multiplicity density
$(dn/d\eta)$ is estimated as twice the density measured in the good-azimuth
region.}. All quantities measured are defined in the notation of correlation 
integrals~\cite{10},\[
r_2 (Q) 
 =  \frac{\rho_2 (Q)}  {\rho_1 \ot \rho_1(Q)}
 =  \frac{\int_{\Omega} d^3 \bp_1 \, d^3 \bp_2 \, \rho_2 (\bp_1,\bp_2) \,
                      \delta \left[Q - q(\bp_1,\bp_2) \right]}  
         {\int_{\Omega} d^3\bp_1\, d^3\bp_2\, \rho_1(\bp_1)\, \rho_1(\bp_2)\,
                      \delta \left[Q - q(\bp_1,\bp_2) \right]}
\]with $\bp$ the three-momenta and $q \equiv \sqrt{-(p_1 - p_2)^2}$,
with $p$ being the corresponding four-momenta. 
The integration region $\Omega$ is identical with our experimental
cuts as specified above. All particles have been assumed to be pions.

In $\bar{p}p$ reactions and in full phase space the number of positive 
and negative particles are equal; furthermore in the central rapidity region 
the corresponding $\rho$-functions are also equal. Given the charged multiplicity
$\rho_1(\bp) = \rho_1^+(\bp) + \rho_1^-(\bp)$,
we hence assume that 
$\rho_1^+(\bp) = \rho_1^-(\bp) \simeq \frac{1}{2} \rho_1(\bp) $ 
and therefore also 
$\rho_1^+ \ot \rho_1^+(Q) \simeq {\textstyle{1\over 4}}\rho_1\ot\rho_1(Q)$
etc. The like-sign and opposite-sign normalised correlation densities hence 
become, respectively,
\begin{eqnarray*}
r_2^\ls (Q) &=& \frac{\rho_2^{\pm\pm} (Q)}{\rho_1^{\pm} \ot 
\rho_1^{\pm} (Q)} = \frac{\rho_2^{++} (Q) + \rho_2^{--}(Q)} {\rho_1^+ 
\ot \rho_1^+ (Q) + \rho_1^- \ot \rho_1^- (Q) } \simeq 
\frac{\rho_2^{++} (Q) + \rho_2^{--}(Q)} { {\textstyle{1\over 2}} 
\rho_1\ot \rho_1(Q) }\,,
\\
r_2^{os} (Q)
&=& \frac{\rho_2^{\pm\mp} (Q)}{\rho_1^{\pm} \ot \rho_1^{\mp} (Q)} = \frac{\rho_2^{+-} (Q) + \rho_2^{-+}(Q)}
        {\rho_1^+ \ot \rho_1^- (Q) + \rho_1^- \ot \rho_1^+ (Q) }
 \simeq \frac{\rho_2^{+-} (Q) + \rho_2^{-+}(Q)}
        { {\textstyle{1\over 2}}  \rho_1\ot \rho_1(Q) } \,.
\end{eqnarray*}
In Fig.\ 1, both the like-sign and opposite-sign correlation densities 
show a strong dependence on multiplicity. To perform, however, a 
quantitative analysis, one has to get rid of the combinatorial 
background by calculating the cumulants and secondly to correct for 
the bias introduced by fixing multiplicity\footnote{
The usual Bose-Einstein analysis assumes that $r_2^\ls$ tends to a constant
for large $Q$, ie.\ $r_{2\ \rm{BE}}^\ls (Q \geq 1)$ = constant. It should be
clear from Figure 1 that no such constancy exists for limited multiplicity
windows.}.

\section{Cumulants for limited multiplicity ranges}

A quantitative study of the bias introduced by fixing multiplicity 
has been performed by Lipa et al.~\cite{8} At fixed total 
multiplicity $N$, the standard factorial 
cumulants
\begin{equation}\label{eqn4}
\kappa_2 (\bp_1, \bp_2\, |N) = \rho_2 (\bp_1,\bp_2\, |N) 
            - \rho_1(\bp_1\, |N) \, \rho_1(\bp_1\, |N)
\end{equation} are nevertheless nonzero, even when particles are completely 
uncorrelated. These purely ``external'' correlations appear 
because the cumulant for any multinomial distribution is nonzero; 
for example, in second order,
\begin{equation}\label{eqn5}
\kappa_2^{\rm{mult}} (\bp_1, \bp_2\, |N) 
= -\frac{1}{N} \rho_1(\bp_1\, |N) \, \rho_1(\bp_2\, |N)\,.
\end{equation} As shown in Ref.~\cite{8}, the ``internal cumulants'' 
\begin{equation}\label{eqn6}
\kappa_2^I (\bp_1, \bp_2\, |N) 
\equiv \kappa_2 (\bp_1,\bp_2 |N) - \kappa_2^{\rm{mult}} (\bp_1, \bp_2\, |N)
\end{equation} correct this bias exactly: they are zero whenever the $N$ particles behave
multinomially. 
Integrating~(\ref{eqn4})--(\ref{eqn6}) to 
obtain the correlation integral and normalizing,  
we arrive at the normalized internal cumulants for given fixed $N$,
\begin{equation}\label{eqn7}
K_2^I (Q \, |N) = \frac{\kappa_2^I (Q \, |N)}{\rho_1 \ot \rho_1(Q \, |N)}  
= r_2 (Q \, |N) - \left(  1 - \frac{1}{N}\right).
\end{equation}
The prescription for calculating the $K_2^I (Q \, |N)$ 
is therefore to measure at fixed $N$ (in $\Omega$) first
$r_2 (Q \, |N)   = {\rho_2 (Q \, |N)} \bigl/ {\rho_1 \ot \rho_1(Q \, |N)}$
and then to subtract
\begin{displaymath}
 1 - \frac{1}{N} = \frac{N(N-1)}{N^2} = 
\frac{\int_{\Omega} \rho_2 (Q \, |N) dQ}{\int_{\Omega} \rho_1 \ot 
\rho_1(Q \, |N) dQ}.
\end{displaymath}
To obtain adequate statistics, we have to measure averages over limited 
multiplicity ranges rather than at fixed $N$. 
The internal second-order cumulant averaged over the multiplicity 
range $[A,B]$ is given in~\cite{8} by
\begin{equation}\label{eqn8}
\overline{\kappa_2}^I (\bp_1,\bp_2) 
= \frac {\sum_{N=A}^B P_N  \kappa_2^I(\bp_1, \bp_2\, |N)} {\sum_{N=A}^B  P_N} 
\,,
\end{equation} where $P_N$ is the experimental multiplicity distribution in $[A,B]$
and the bar over any quantity $S$ denotes
$\overline{S} \equiv \sum_{N=A}^B P_N S(N) / \sum_{N=A}^B P_N$.

\noindent Assuming that the  shape of $\rho_1(\bp\, |N)$   does not vary in $[A,B]$
\begin{displaymath}
\rho_1(\bp\, |N)   \simeq 
\left( N / \overline{N}\right)  \overline{\rho_1} (\bp_1) \,,
\end{displaymath} and adopting again the correlation integral prescription, we can write
\begin{equation}\label{eqn9}
\overline{K_2}^I (Q) = \frac{\overline{\kappa_2}^I (Q)}
                       {\overline{\rho}_1 \ot \overline{\rho}_1 (Q)}    
= 
  \frac{\overline{\rho_2} (Q)}
       {\overline{\rho}_1 \ot \overline{\rho}_1 (Q)} 
- \frac{\int_{\Omega} \overline{\rho}_2 (Q) dQ}
       {\int_{\Omega} \overline{\rho}_1\ot\overline{\rho}_1(Q) dQ}\,.
\end{equation}

\begin{figure}
\psfig{figure=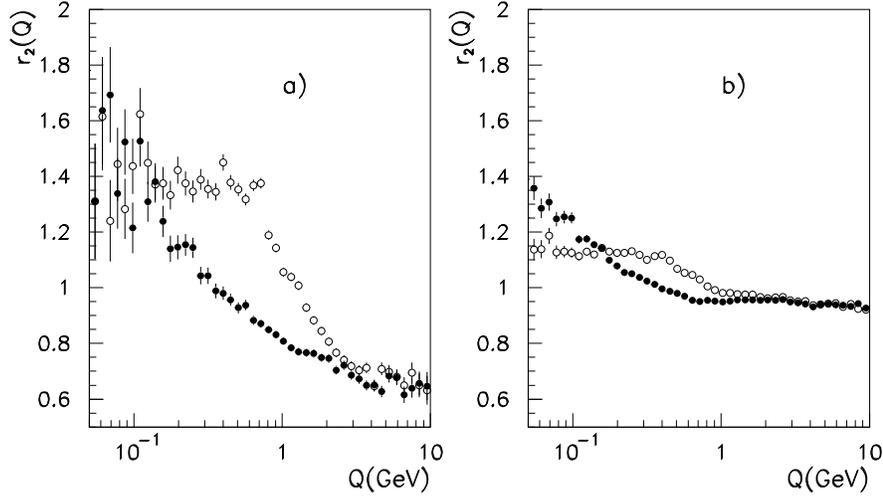,height=3.0in}
\vspace{-1.1cm}
\caption{$r_2^\ls$ versus $Q$ for like-sign pairs (full 
circles) and $r_2^{os}$ versus $Q$ for opposite-sign pairs (open 
cirles) at a) $dn/d\eta = 0.83$ and b) $dn/d\eta = 5.3$.}
\end{figure} 
\vspace{1cm}

\begin{figure}
\psfig{figure=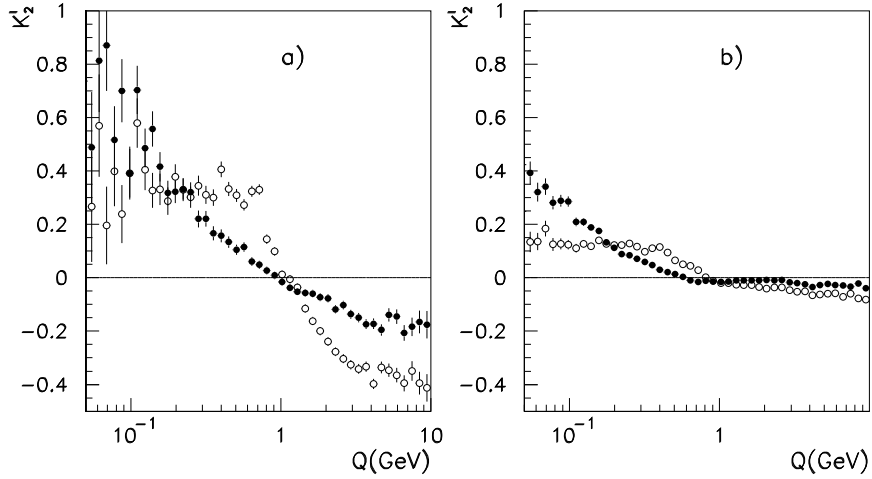,height=3.0in}
\vspace{-1cm}
\caption{Internal cumulants eqns.~(\ref{eqn10}) for 
$\ell s$ pairs (full circles) and $os$ pairs (open circles) for two
different multiplicity densities, a) $dn/d\eta = 0.83$ and b) 
$dn/d\eta = 5.3$.}
\end{figure}

\noindent For like-sign measurements, this becomes
\begin{equation}\label{eqn10}
\overline{K_2}^{I\, \ls}   
=  \frac{\overline{\rho}_2^\ls (Q)}
        { \textstyle{1\over 2} \overline{\rho}_1\ot\overline{\rho}_1 (Q)} 
- \frac{\int_{\Omega} \overline{\rho}_2^\ls (Q) dQ}
         {\textstyle{1\over 2} \int_{\Omega} 
                               \overline{\rho}_1\ot\overline{\rho}_1(Q) dQ}
\end{equation} and similarly for $\overline{K_2}^{\rm{I}\,os}$.
Since all quantities shown here and below are to be 
understood as mean values in multiplicity ranges like in 
(\ref{eqn8}) -- (\ref{eqn10}), we henceforth 
(and in Fig.\ 1) omit the bar on the symbols.

Fig.\ 2 shows both like- and opposite-sign internal cumulants 
for two selections of $dn/d\eta$. Three features are immediately
apparent: 
\begin{itemize}

\item[I.]
The cumulants differ from the moments of Fig.\ 1 by a shift constant
in $Q$, but the amount of the shift changes for different cumulants.
The importance of changing from $r_2$ to $K_2^I$ lies in the 
fact that the latter demarcate clearly the ``line of no correlation''.

\item[II.]
Internal cumulants integrate to zero over the entire phase space;
hence the positive part of $K_2^I$ at small $Q$ is compensated by 
a negative part at larger $Q$. Physically, this means that particles like
to cluster, so that there is a surfeit of pairs at small $Q$ and a
dearth of pairs at large $Q$ compared to the uncorrelated case.

\item[III.]
The dependencies of the like- and opposite-sign cumulants on
multiplicity are rather similar in that both decrease markedly with 
$dn/d\eta$. This is discussed below.
\end{itemize}

\section{Multiplicity dependence of normalized cumulants}
The similar decrease of $K_2^{I\, \ls}$ and $K_2^{I\, os}$ with $dn/d\eta$
suggests that both could have the same functional dependence on multiplicity 
density. Under this hypothesis,
\begin{eqnarray}\label{eqn12}
K_2^{I\, \ls} (Q, dn/d\eta ) &=& Y^\ls(Q)  C (dn/d\eta , Q) \,,  \\
K_2^{I\, os} (Q, dn/d\eta ) &=& Y^{os} (Q)  C (dn/d\eta , Q) \,, \nonumber
\end{eqnarray} the quotient of the cumulants should be independent of multiplicity
\begin{equation}\label{eqn13}
\frac{K_2^{I\, \ls} (Q, dn/d\eta )}{K_2^{I\, os} (Q, 
dn/d\eta )} \quad = \quad \frac{Y^\ls(Q)}{Y^{os} (Q)}
\ \ =  \ \ 
\left( \mbox{constant in\ } \frac{dn}{d\eta} \right) \,.
\end{equation}
Figure 3a shows that (\ref{eqn13}) holds at least approximately 
within error bars. (In the region where both cumulants are near zero,
no meaningful quotients can be formed.)

Having shown that like- and unlike-sign internal cumulants
behave approximately in the same way as functions of $dn/d\eta$,
we now ask how this dependence is structured. A first hypothesis,
is that $K_2^I$ depends inversely on $N$;
in the notation of Eq.\ (\ref{eqn12}),

\begin{equation} \label{eqnff}
K_2^{I\, a}(Q\, |N) = Y^a(Q) C(N, Q) = Y^a(Q) N^{-1} \qquad  a = \ls, os \,.
\end{equation}

This can be motivated theoretically by
\begin{itemize}
\item[a)] 
{\bf Resonances:}
If the unnormalized cumulants 
$\kappa_2^{I\, os}$ and $\kappa_2^{I\, \ls}$ were wholly the result of 
resonance decays and if the number of resonances were 
proportional to the multiplicity $N$, then $\kappa_2^I \propto N$.
Assuming $\rho_1(\bp\, |N) \propto N \rho_1(\bp)$ 
 gives $\rho_1 \ot \rho_1 \propto N^2$, and hence after normalization,
the resonance-inspired guess 
\[
K_2^I = \frac{\kappa_2^{I\, {\rm res}}}  {\rho_1\ot\rho_1}
 \propto \frac{1}{N} \,.
\]
\item[b)] 
{\bf Independent superposition} in momentum space of $n$ equal 
strings would also lead to
$K_2^I = n \cdot \kappa_2^{I\, {\rm string}} / (\rho_1\ot\rho_1)
\propto (1/n) \propto (1/N)$.
Deviations can occur if the strings are unequal.
\end{itemize}
Eq.\ (\ref{eqnff}) implies that
$K_2^{Ia}(Q\, |N_1)/K_2^{Ia}(Q\, |N_2) = (N_2/N_1) =$ (constant in $Q$)
for two multiplicities $N_1$ and $N_2$.
In Fig.\ 3b, we show the quotient of cumulants for two multiplicities.
Surprisingly, we find not one but two constants, one for small
$Q \lsim 0.4$~GeV, one for large $Q \gsim 2$~GeV, where only the
latter corresponds to the value $(N_2/N_1)$ expected from 
(\ref{eqnff}), shown as the
dotted line. At small $Q$, one must clearly look for other functional 
forms for $C(N,Q)$. Some phenomenological guesses are as follows.~\cite{11} 

\begin{itemize}

\item[c)] 
In the {\bf source picture of Bose-Einstein} correlations, 
the independent superposition of sources in configuration space
leads to
$K_2^{I, BE} \simeq$ constant in $N$. We would expect $\lambda$ to be 
independent of multiplicity, but the radius $R$ to increase with $N$.
Hence there should be a change of shape of $K_2^{I\, \ls} (Q)$ 
but no overall $N$-dependence.

\item[d)] 
A mixture of processes, a) with c) could result in a dependence

\begin{equation}\label{eqn16}
K_2^{I\, \ls} (Q\, |N) \approx a(Q) + \frac{b(Q)}{N}
\end{equation}

However, no comparable source picture is available for the unlike-sign
case. Also, the quotient
$K_2^{I\, \ls}/K_2^{I\, os}$ would not be constant in $N$, in
contradiction to the results of Fig.\ 3a. 

\item[e)]
Another guess would yield a $N^{-1}$ dependence for large $Q$, 
while the small-$Q$ region 
scales with $N^{-\alpha}$, with a different ``constant'' $Y_2$,
\begin{equation}  \label{eqnggg}
K_2^{I\, a} = \frac {Y_1^a(Q)} {N} \Theta(Q-2)
            + \frac {Y_2^a(Q)} {N^\alpha} \Theta(0.4-Q) + \ldots 
\qquad a = \ls, os\,,
\end{equation}
\end{itemize}
\vspace{-4mm}

\begin{figure}[h]
\psfig{figure=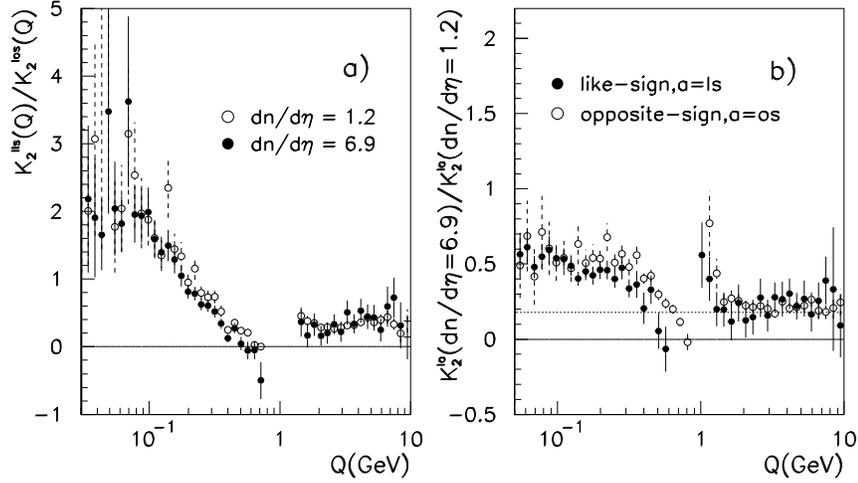,height=3.0in}
\vspace{-1cm}
\caption{a) The ratio Eq.\ (\ref{eqn13}) is shown for two selections of 
$dn/d\eta$.  b) The ratio of two $K^{I\, a}_{2} (Q)$ at different 
$dn/d\eta$ as indicated on ordinate. The 
region around $Q=1$ has been omitted because the denominators are around 
zero.}
\end{figure} 

\vspace{0.3cm}

In order to test the above ideas, we plot the cumulants against 
$(dn/d\eta)^{-1}$ as follows.  To avoid local statistical 
fluctuations, the normalized cumulants $K_2^{I\, \ls} (Q)$ and 
$K_2^{I\, os} (Q)$ are fitted with suitable functions in restricted 
$Q$-ranges (not shown).  
The $K_2^{I\, \ls}(Q)$ have also been fitted to an exponential 
parametrization for $Q < 1$~GeV/c \begin{equation}\label{eqn14}
K_2^{I\, \ls} (Q) = a + \lambda \ e^{-RQ}\,.
\end{equation}
The multiplicity dependence of the fitted $K_2^{I\, \ls}$ and 
$K_2^{I\, os}$ (circles), 
as well as the corresponding $\lambda$ values 
(crosses) are plotted in Fig.\ 4.
Fig.\ 4b shows that, as in Fig.\ 3b, the $(1/N)$-dependence is satisfied 
only for large $Q$, but not for small $Q$. 
The $a + b/N$ dependence in Fig.~4a (solid line) provides a possible,
but hardly unique, explanation. 
The dashed line corresponding to $N^{-\alpha}$
behaviour (\ref{eqnggg}) seems somewhat better. However, while
this form is compatible with all results shown so far, we have
no phenomenological justification for it.
Eq.~(\ref{eqnggg}) also omits the region around $Q\simeq 1$~GeV (e.g.\ the 
region of $\rho^{0}$ production) where the phase space contributes 
maximally, but the $K^{I\,os}_{2}$ decrease rapidly with increasing 
$Q$ and the $K^{I\, \ell s}_{2}$ are already small (Fig.~2). It is 
difficult to investigate this important region directly. 
A $1/N$ dependence (via resonances, for example) in the dominant region 
around 1 GeV/c could presumably cause the large-$Q$ region to follow
suit via missing pairs.

\begin{figure}[h]
\psfig{figure=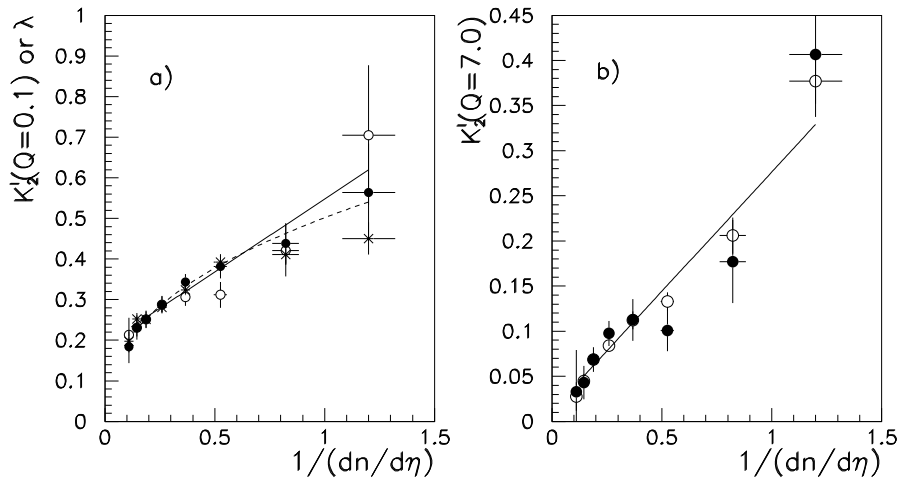,height=3.0in}
\vspace{-1cm}
\caption{a) Multiplicity dependence of $K_2^{I\, \ls} (Q=0.1)$ 
(full circles) and of $K_2^{I\, os} (Q=0.1)$ (open circles) and 
$\lambda$ values of exponential fit, Eq.\ (\ref{eqn14}) (crosses), b) 
as in a) but for the large Q region. 
For better comparison of the respective dependencies on $dn/d\eta$, the 
absolute values have been scaled by constant factors.
The straight lines are linear fits, the dashed line corresponds to 
eqn.(\ref{eqnggg}).}
\end{figure}

\section{Summary}

The multiplicity dependence of like-sign and opposite-sign two-body 
correlation functions are studied with the same model-independent 
strategy. The bias introduced by selecting events of a given overall 
multiplicity is eliminated in measuring  ``internal  cumulants''.
We observe that 
\begin{itemize}
\item[-] the like-sign and opposite-sign functions have a very 
similar multiplicity dependence, which is surprising because of their 
different shapes and assumed origins,
\item[-] there exist two regions, one at small $Q$, where the multiplicity 
dependence of both is smaller than $1/N$, and one at large $Q (\gsim 
1$~GeV/c), where the functions are negative and follow roughly an 
$1/N$ law.
\end{itemize}

Theoretical work\cite{5,6,13} is challenged by these findings.  The 
similar behaviour of $\ell s$ and $os$ functions at small $Q$, both 
decreasing with multiplicity, favours suppression of Bose-Einstein 
correlations between products of different strings.  Resonance decays 
(Regge terms) should arguably contribute also to 
$\ls$-functions~\cite{11,12}.  One could therefore try to explain 
alternatively the decrease of $\ls$ functions with $N$ by a mixture of 
Bose-Einstein correlations (assumed to be constant in $N$) with 
resonance production.  This leaves unexplained, however, the similar 
behaviour of $os$ and $\ell s$ functions and the nearly $N$ independent 
shape\footnote{The reported increase~\cite{2,3,4} of $R$ with $N$ is 
dependent on the choice of the fit function and region.  Excluding the 
region $Q
\geq 1$~GeV/c we observe at small $Q$ only a minimal or even missing 
increase of $R$ (from fits to (\ref{eqn14})) with $N$.} in the 
small-$Q$ region.

\section*{Acknowledgements}
We thank T.\ Cs\"org\"o and his team for their kind hospitality and
support.
This work was funded in part by the Foundation for Research Development.


\section*{References}
\begin{thebibliography}{99}

\bibitem{1} A.\ De Angelis, Proc.  27th.\ Int.\ Symp.\ on 
Multiparticle Dynamics, Frascati, Italy, Sept.\ 1997.

\bibitem{5} B.\ Andersson and W.\ Hofmann, {\it Phys.\ Lett.\/} B{\bf 169},364 (1986).

\bibitem{6} B.\ Andersson and M.\ Ringn\'er, 
            preprint LUTP 97--07, {\it hep-ph/9704383};
            J.\ H\"akkinen and M. Ringn\'er, 
            preprint LUTP 97--32.

\bibitem{7} L.\ L\"onnblad and T.\ Sj\"ostrand, 
            preprint LUTP 97--3, {\it hep--ph/9711460}.

\bibitem{8} P.\ Lipa, H.C.\ Eggers, and B.\ Buschbeck, 
            {\it Phys.\ Rev.\/} D{\bf 53}, R4711 (1996).

\bibitem{9} C.\ Albajar et al.\ (UA1), 
            {\it Z.\ Phys.\/} C{\bf 44}, 15 (1989);\\
            M.\ Calvetti et al., 
            {\it IEEE Trans. Nucl. Science NS--30}, 71 (1983).

\bibitem{10} P.\ Lipa et al., 
            {\it Phys.\ Lett.\/} B{\bf 285}, 300 (1992);\\
            H.C.\ Eggers et al., 
            {\it Phys.\ Lett.\/} B{\bf 301}, 298 (1993);\\
            H.C.\ Eggers et al., 
            {\it Phys.\ Rev.\/} D{\bf 48}, 2040 (1993).

\bibitem{11} We thank A.\ Bia\l as and K.\ Fia\l kowski for  
            discussions on this topic.

\bibitem{2} C.\ Albajar et al.\ (UA1), 
            {\it Phys.\ Lett.\/} B{\bf 226}, 410 (1989).
\bibitem{3} 
            T.\ \AA kesson et al. (AFS), 
            {\it Phys.\ Lett.\/} B{\bf 187}, 420 (1987).
\bibitem{4} 
            A.\ Breakstone et al. (SFM), 
            {\it Z.\ Phys.\/} C{\bf 33}, 333 (1987).
            
            
            \bibitem{13} N. Suzuki, this conference.
            
\bibitem{12} 
             E.L.\ Berger et al.,  
            {\it Phys.\ Rev.\/} D{\bf 15}, 206 (1977).
            
            
            
            
            

\end{thebibliography}
\end{document}